\documentclass{IOS-Book-Article}

\usepackage{mathptmx}
\usepackage{graphicx}
\usepackage{soul}\setuldepth{article}
\usepackage{times}
\normalfont
\usepackage[T1]{fontenc}

\def\hb{\hbox to 11.5 cm{}}
   
\begin{document}
\pagestyle{headings}
\def\thepage{}
\begin{frontmatter}              

\title{Towards Open Diversity-Aware\\ Social Interactions}


\author[A,B]{\fnms{Loizos} \snm{Michael}}
\author[C]{\fnms{ Ivano} \snm{Bison}}
\author[C]{\fnms{ Matteo} \snm{Busso}}
\author[D]{\fnms{ Luca} \snm{Cernuzzi}}
\author[E]{\fnms{ Amalia} \snm{De G\"otzen}}
\newline
\author[F]{\fnms{Shyam} \snm{Diwakar}}
\author[G]{\fnms{ Kobi} \snm{Gal}}
\author[H]{\fnms{ Amarsanaa} \snm{Ganbold}}
\author[I]{\fnms{ George} \snm{Gaskell}}
\newline
\author[J]{\fnms{Daniel} \snm{Gatica-Perez}}
\author[K]{\fnms{ Jessica} \snm{Heesen}}
\author[L]{\fnms{ Daniele} \snm{Miorandi}}
\author[M]{\fnms{ Nardine} \snm{Osman}}
\newline
\author[N]{\fnms{Salvador} \snm{Ruiz-Correa}}
\author[K]{\fnms{ Laura} \snm{Schelenz}}
\author[G]{\fnms{ Avi} \snm{Segal}}
\author[M]{\fnms{ Carles} \snm{Sierra}}
\newline
\author[O]{\fnms{Hao} \snm{Xu}}
\author[C]{\fnms{ Fausto} \snm{Giunchiglia}
\thanks{Corresponding Author: Fausto Giunchiglia, fausto.giunchiglia@unitn.it.}}

\runningauthor{}
\address[1]{Open University of Cyprus, Latsia, Cyprus}
\address[2]{CYENS Center of Excellence, Nicosia, Cyprus}
\address[3]{University of Trento, Italy}
\address[4]{Universidad Católica ``Nuestra Señora de la Asunción'', Asuncion, Paraguay}
\address[5]{Aalborg University, Copenaghen, Denmark}
\address[6]{Amrita Vishwa Vidyapeetham, Amritapuri campus, Kerala, India}
\address[7]{Ben-Gurion University, Beer-Sheva, Israel}
\address[8]{National University of Mongolia, Ulanbataar, Mongolia}
\address[9]{London School of Economics and Political Science, London, U.K.}
\address[10]{Idiap Research Institute and EPFL, Switzerland }
\address[11]{University of Tuebingen, Tuebingen, Germany}
\address[12]{U-Hopper, Trento, Italy}
\address[13]{IIIA-CSIC, Barcelona, Spain}
\address[14]{IPICYT, Mexico}
\address[15]{Jilin University, Changchun, China}

\begin{abstract}
Social Media and the Internet have catalyzed an unprecedented potential for exposure to human diversity in terms of demographics, talents, opinions, knowledge, and the like. However, this potential has not come with new, much needed, instruments and skills to harness it. This paper presents our work on promoting richer and deeper social relations through the design and development of the ``\emph{Internet of Us}'', an online platform that uses diversity-aware Artificial Intelligence to mediate and empower human social interactions. We discuss the multiple facets of diversity in social settings, the multidisciplinary work that is required to reap the benefits of diversity, and the vision for a diversity-aware hybrid human-AI society.
\end{abstract}

\end{frontmatter}


\section{Introduction}

No person is an island. No one knows or is skilled in everything. Human societies empower us to reach out to people with skills and information complementary to ours; we rely on others, and they rely on us. Diversity --- our ``differences'' in language \cite{UKC-IJCAI,giunchiglia2023representing,koch2024layers,bella2024tackling}, knowledge \cite{KD-2020-Giunchiglia-KR,giunchiglia2022representation}, routines, social practices, competencies, and skills \cite{schelenz2021theory,mercado2023social,girardini2023adaptation} --- is a desired feature that should be treasured and sought after. But, as social interactions increasingly shift from the physical to the virtual world, where ``the same'' is reinforced via biased algorithmic recommendations \cite{orphanou2022mitigating}, and as societies consequently become more fragmented, this diversity in the access we have to information and social interactions is inevitably restricted \cite{cinelli2021echo}, and at worst, is viewed as an undesirable bug to be avoided and eliminated.

Not unlike the \emph{Internet of Things}, a communication platform for the interaction of diverse devices, this paper puts forward the vision of an ``\emph{Internet of Us}'', a hybrid human-AI platform, built on top of the \textit{iLog} data collection APP \cite{zeni2014multi}, over which diverse communities and individuals can co-exist and interact \cite{KD-2020-Osman}. The platform  enables social interactions that reach beyond the limited social sphere of each individual, empowers humans to explore more fully the \emph{unknown unknowns} of their social search space, and helps them capitalize on the range of their diverse collective interests and competencies. Acting as an ecosystem, the platform supports \emph{app developers} by providing them with functionalities to implement diversity-aware apps (and leaving them to focus on the app logic only, and on fulfilling the needs of potential users), \emph{end users} in their online social interactions through the use of the developed applications, and \emph{researchers} by providing them with social interaction data for analysis purposes.

The platform has been used and tested in several pilot studies \cite{KD-2022-WeNetDiversityOnePlus,Kun}, engaging with thousands of students from across the world\footnote{Participating students were attending the following: Jilin University (China), Amrita Vishwa Vidyapeetham (India), Aalborg University (Denmark), London School of Economics and Political Science (U.K.), University of Trento (Italy), National University of Mongolia (Mongolia), Universidad Católica ``Nuestra Señora de la Asunción'' (Paraguay), and Institute for Scientific and Technological Research of San Luis Potosi (Mexico).} over the span of four years, and exploring the benefits that a diversity-aware chat application could bring to people, and how people interpret diversity when being presented with it in social interaction settings. Our empirical investigation utilized two different typologies of pilots. The so called ``\emph{diversity pilots}'' sought to collect data about the students’ characteristics and behaviors through online surveys and mobile phone sensors towards refining the model of diversity implemented in the platform. The so called ``\emph{application pilots}'' sought to explore the needs and interactions of the students in real-life scenarios, leveraging their diversity through the integration in the platform of research-driven technological solutions. Data collected through the diversity pilots was used to develop and test measurement instruments on human behavior, and to study diversity from a sociological perspective, while also informing the structure of participant profiles in the platform. At the same time, the specific information provided by the users was used to populate their profiles when they participated in the application pilots, and to support research on developing machine learning and interaction algorithms that utilize diversity in the behavioral routines of users.

We organize the contributions of this paper as follows. First, we present the results of a series of experiments in the wild whose goal is to understand a set of crucial dimensions which can be used to model the diversity of people (Section \ref{S2}, which presents the results from the diversity pilots) as well as the needs of people seeking support (Section \ref{S3}, which reports the results from the application pilots). Then, we discuss the key mechanisms that enable and incentivize the creation of communities (Section \ref{S4}). We continue to discuss the issues that relate to the modeling and learning about the personal and social context of people (Section \ref{S5}). Finally, we elaborate a few considerations which show how ethics can be instrumental towards the establishment of the ``Internet of Us'' (Section \ref{S6}). 


\section{Understanding and Modeling (the Characteristics of) Human Diversity}
\label{S2}
 
A key prerequisite for achieving the goal of our platform is the ability to capture diversity within a community and to represent it in a machine-readable fashion. We have identified two main classes of diversity dimensions across which diversity is measured \cite{Tsui1991,Tsui1992,Harrison1998}: 

\begin{itemize}
    \item \emph{visible} (or ``observable'') demographic characteristics, e.g., sex and gender, skin color / ethnicity, physical (dis)abilities, and age; and 
    \item \emph{invisible} (or ``deep'') characteristics, e.g., ethnic / cultural / socioeconomic background, personality characteristics, cognitive and interaction attitudes, social / human values, competencies and education, type of work organization, and role. 
\end{itemize}

\noindent
The complete list of diversity dimensions, together with a description of the experiments aimed at collecting these data, as well as of the datasets that were collected, is provided in \cite{KD-2022-WeNetDiversityOnePlus}. Among the many participant dimensions that we have modeled, a few key invisible characteristics can be grouped into three abstract dimensions, usually considered as part of the general theory of the \emph{Communities of Practice}: \emph{materiality}, \emph{competencies}, and \emph{meaning} \cite{shove2005, Rapke2009}. Materiality relates to the tangible assets that an individual has access to. Competencies incorporate an individual's endowed skills, (background) knowledge, as well as social and relational skills required to perform a practice. Meaning incorporates understandings, beliefs, values, lifestyle, emotions, and social and symbolic significances, which are not transferable but can be learned through socialization \cite{berger1996}. Meaning makes it possible to connect materiality and competencies to give rise to ``behavioral routine" actions that become practices at a social level. It also acts as a ``nexus'', not only within a single practice, but in linking different practices, and even in creating new ones.

Both the visible and the invisible characteristics are operationalised in the profile of each community member. The vector of values of these fields associates, then, each individual with a point in a multi-dimensional space, which is to be searched when seeking to establish social relations between community members. Imposing a distance metric over this search space allows a notion of variability between individuals to be formalized. Notably, however, when developing concrete apps for particular communities over the platform, some of these characteristics may become irrelevant. For instance, when asking for a certain piece of information, the ``materiality'' dimension tends to become less important in determining an appropriate individual, or might need to be appropriately ``interpreted''. The general approach is that the platform provides support for all the diversity dimensions that we have identified, thus letting the different apps developed on top of it the decision of which to use. The work in \cite{10.1145/3569483} describes some initial studies of the explicative power of the diversity dimensions that we have selected to model.

The richness of the profiles comes from a descriptive view of diversity, which captures the variability within a community (or a subset chosen to receive a request), but also between the person seeking support and each potential respondent to that request. Being able to compute degrees for these two types of diversity is a step towards making the members of a community aware of the diversity that exists within their ranks and files. A community member seeking support can then explicitly make use of this awareness by identifying which dimensions of the search space to restrict, and in what way, so that the exploration of the search space can become feasible, but not overly restrictive.

\section{Understanding the Needs and Constraints of Humans Seeking Support}
\label{S3}

To study the perception of users when presented with diversity, we developed a technology probe \cite{Hutchinson} in the form of a Q\&A chat app with a lightweight interface (avoiding the technical complexity of a full-fledged conversational agent with natural language processing capabilities) that connects people in a community. In the application pilots, users could pose a support query in written text, select the domain of their query (hobbies, arts, career, studies), specify restrictions on who should view their query, and note whether the query was of a sensitive nature and should be submitted anonymously. This information was used to construct a ranking of users, who were then asked to respond to the query. 

The chat app was evaluated favorably in the pilots and was appreciated for its ability of connecting students from different departments / education and cultural backgrounds, and with different skills and competencies. While the possibility of getting support from ``expert users'' was an acknowledged feature of the app, a major attraction for some users was contacting other students, making social connections, and creating the feeling of a community while exploring its diversity. Most of the expectations towards diversity carried an exploratory quality, where people were driven by curiosity, or wished serendipity in posing queries to people different from themselves. This need was particularly apparent in the pilots ran during the COVID-19 pandemic. Exit surveys and focus groups with the student users following the most recent pilots reaffirmed the finding that many users actively sought to capitalize on diversity, and showed that giving advice was marginally more satisfying than receiving it, that the absence of photographs and the opportunity to participate anonymously was welcomed, and that flagging up a question as sensitive was appreciated and taken to reflect a positive ethic that guided the design of the chat app. 

Beyond the aforementioned pilot studies and our developed Q\&A chat app, the platform was also utilized by a number of external third parties who were interested in developing alternative use cases using our technology. One such external third party was the Spanish NGO Fundaci\'on Cibervoluntarios, which coordinates a network of more than 1,800 cybervolunteers helping citizens to level out digital gaps. In March 2020, during the lockdown caused by the COVID-19 pandemic, Fundaci\'on Cibervoluntarios set up a helpline to provide support to citizens having doubts or problems related to technology, computing devices, and their connectivity, configuration, usage, or cybersecurity. Fundaci\'on Cibervoluntarios piloted our technology to empower cybervolunteers to effectively help out citizens in difficulty. Another external third party was the Greek social cooperative enterprise CommonsLab, which leveraged our technology to implement a chat application (MaTSE: a Matchmaking Tool for Social Entrepreneurs) that brings together different constituencies in the social innovation landscape: entrepreneurs with a business idea for a social innovation venture, potential investors, and other actors who may help in the process (e.g., established entrepreneurs who could provide early feedback, etc.). 

As part of our effort to utilize (and evaluate) our technology in diverse contexts, we have been involved in two additional studies. The first study sought to elicit the perspective of Afghan refugees in Germany on the Q\&A chat app. Afghan refugees are particularly marginalized technology users because of their experiences of political, economic, and social violence in Afghanistan and abroad. The study span four months in 2022, and was conducted in cooperation with an NGO for the rights of Afghan women. In conversations with 14 Afghan refugee women, it became clear that empowerment through technology is not inevitable. The Q\&A chat app presented significant barriers for their participation due to the ``complicated'' registration process and usage commands. 

The second study sought to apply our technology as a way to alleviate the diminished social interactions between the Greek-Cypriot and Turkish-Cypriot communities of Cyprus, to the extent these interactions are affected by the language barrier. A simplified version of the Q\&A chat app was extended with the extra functionality of allowing each user to interact (with the app and others) in their native language, while supporting the automatic translation of messages exchanged between the two communities. The effort itself has been met with interest from members of both communities, but has also highlighted the need to be wary of unforeseen pitfalls, such as the potential of mistranslation that may be deemed offensive by the recipient of the message. The two studies have highlighted that an ``Internet of Us'' must consider diversity in digital literacy and language skills, and ensure that the app design is inclusive of low-literate and socio-economically marginalized groups, and non-homogeneous groups that might not exhibit a shared trust.


\section{Supporting and Incentivizing the Self-Determination of Communities}
\label{S4}

Awareness of descriptive diversity and the dimensions along which it is measured, supports the self-determination of communities by providing the primitive concepts against which a community can differentiate itself from other communities. A community could, for instance, self-determine as including only residents of Spain who are wine aficionados, effectively utilizing the ``country of residence'' and the ``hobbies and leasure'' features of its members as the properties that make it distinct from other communities.

The same primitive concepts support a second path to community self-determination, by providing the basis over which a community can define the \emph{norms} that apply during the interaction between its members. Sociologically speaking, norms capture the rules of engagement, or the social rules of behavior of a community and its members, and are the direct application of the social values of that community. They capture community-wide provisions (e.g., do not ask the same person for support more than three times within any single week), or individual provisions (e.g., do not share a request for support with members blocked by an individual). Norms are also meant to interpret the search directives provided by members of a community. The same cue (e.g., seek support from people nearby) can be interpreted differently if seeking support within one's ``neighborhood watch'' community, or within one's ``alma mater graduates'' community. Technologically speaking, norms define the protocol of exploring the search space when seeking to solve the search problem of identifying appropriate respondents to a sequence of queries.

Norms come in the form of natural language expressions in human societies, and, traditionally, in the form of logic-based expressions in formal machine-readable representations (e.g., constraint-based \cite{Garcia-CaminoRSV09}, time-based \cite{AgotnesHRSW09}, or event-based \cite{KowalskiS05}). In our work, we have adopted a declarative representation of norms through conditional logic-based rules~\cite{OsmanCSSG21}. The rule premises refer to profile features (such as gender, competencies, etc.) or other state properties of the platform (such as the number of active community members, the total number of queries, the level of user satisfaction, etc.) that can encode information about a community, a user, or even a task or interaction. The rule conclusions refer to the possible actions that can be performed, which include message-passing actions (such as sending a message to another user, or sending a message to an app's user-facing interface to enable a given button), and actions that change profile features or state properties (such as the privacy setting of one’s location, or the rating for a given task). A norm engine is then designed to mediate interactions by checking their compliance with the relevant norms in place, be they community-wide norms or individual norms. 

Of course, having community members aligned in their understanding of norms is crucial, but difficult in open systems. For instance, if hate speech is prohibited by a norm, not all members might have the same understanding of what constitutes hate speech. We have developed learning mechanisms that can learn the meaning of a norm from a community’s interactions, explain to community members that learned meaning, and adapt over time what is learned through members’ feedback~\cite{SantosOS22}. Furthermore, working with norms allows us to formally analyze the outcomes of interactions~\cite{MontesOS22}. For instance, we can analyze whether certain norms promote / demote certain human values~\cite{abs-2110-09240}, as well as synthesize norms to ensure the maximum promotion of given human values~\cite{MontesS21}. 

How can we promote the growth of active, inclusive, and diverse communities, taking also into account that the platform allows for interactions with \emph{unknown unknowns}? In the spirit of community self-determination, the key consideration is that communities may not wish for external guidelines on inclusiveness and active participation to be applied in a paternalizing and all-knowing manner. Consequently, instead of imposing such guidelines, our platform seeks to incentivize their adoption, taking into account the diverse characteristics of different participants and communities, and avoiding a ``one-size-fits-all'' approach common in current systems. Past works have recognized user diversity and harnessed it for generating incentives using badges~\cite{yanovsky2019one}, text messages~\cite{segal2018optimizing}, and recommended items~\cite{ben2022generating}. Recently, we proposed a diversity-aware multi-armed bandit approach \cite{botta2023sequencing,botta2023personalizing}, which reasons about users' diversity characteristics when making action recommendations. Such reasoning was shown to lead to improved outcomes compared to a ``static'' approach which ignored participants' rich and changing profiles (cf.\ Section~\ref{S2}). 

The platform's incentivization mechanism directs participants to engage with others by seeking support, responding to support queries, and rating the responses they receive, towards becoming more active members of their community. This same mechanism could be used to incentivize participants to not apply search filters unnecessarily, and to actively request for the recipients of their query to be diversified (cf.\ Section~\ref{S5}). 


\section{Modeling the Personal and the Social Context of Individual Participants}
\label{S5}

One of the aspects that makes individuals diverse, and that affects their social interactions through the application of norms, is their personal context \cite{KD-2017-PERCOM,bontempelli2022lifelong}. Due to its inherently dynamic nature, the personal context of any given platform participant cannot be populated in their profile when they first join the platform, but needs to be learned. The world-wide deployment of diversity pilots over our platform presented a unique opportunity to investigate how personal contexts can be learned across diverse communities. Short surveys about the participants' current place, activities, social situation, and mood were collected in parallel with data from smartphone sensors and mobile app logs. The datasets were used to investigate how diversity attributes like the country of residence of participants can be utilized to build ML-based inference models from mobile data. Such an investigation allows for the systematic comparison of country-specific and country-agnostic models for tasks like activity recognition \cite{Bouton22} or mood recognition \cite{10.1145/3569483}, and facilitates the empirical evaluation of whether model transfer is possible across communities. 

The results show that taking country-level diversity into account as part of model learning is important from the perspective of recognition performance, and that transfer of models across countries is not a straightforward task. In this sense, the results contribute towards highlighting the fundamental need for locally-valid data that represents the reality of the world beyond economically advantaged regions \cite{GaticaPerez19}, and for documenting the resulting models with respect to these issues \cite{10.1145/3458723}. More broadly, our empirical results emphasize the need for machine learning models to be aware of the diversity that exists between communities, and by extension, the need for this diversity to be reflected in the learning data that is available (in the participant profiles, in the case of our work). 
  
Beyond their personal context, a profile also includes an individual's social context, which in its simplest form comprises a list of learnable social tie strengths \cite{Perikos.2022} that capture the frequency and manner in which the individual has interacted with other platform participants. A more advanced form of social context comes through a set of norms that specify an individual's policy on whether / how to interact with other participants. This policy can be thought to be more persistent than an individual's social tie strengths, but can still be revised over time. Our work has focused on developing a formal declarative policy-representation language \cite{Markos.2022.prudens} that acknowledges the key role of argumentation as a calculus for human-centric AI \cite{Dietz.2022.calculus}. The language is accompanied by a policy elicitation paradigm \cite{Michael.2019.coaching,Michael.2023.coachable} that is cognitively light and computationally efficient \cite{Markos.2022.proxy}. To support the utilization of this formal policy language, we have investigated how policies can be elicited through natural language cues \cite{Ioannou.2021.translation}, and how they can be integrated with machine learning models \cite{Tsamoura.2021}, towards making them explainable \cite{Michael.2023.fourthAI} and contestable \cite{Tubella.2020}.

The use of individual norms to condition a user's interactions on their personal and social context, and the dependence of the user's context on their interactions may lead to a vicious spiral of diversity-reducing interactions. To compensate for this possibility, we adopt a second view of diversity as a prescriptive value to be promoted, and the platform diversifies the ranking of participants that receive any given query. One might think that diversification should be applied maximally on all visible characteristics, but not on any of the invisible ones, which relate to the abilities of participants to respond to queries and should, therefore, be allowed to be constrained. This view is understandable, given that human societies have practically equated the notion of diversity with that of inclusiveness based on visible characteristics. But this, by itself, does not make this a valid view. 

In certain scenarios, one may wish to select participants based on their visible characteristics, so as to target, for instance, a certain minority demographic group. In other scenarios, one may wish to diversify based on the participants' invisible characteristics, not necessarily as a way to avoid discrimination, but as a way to get a more varied perspective (or serendipity, cf.\ Section~\ref{S3}). In the spirit of self-determination, our diversification process allows each community to choose the profile characteristics (one or many, visible or invisible) against which to diversify, and the extent to which the diversity requirement will override a user's own preferences \cite{Markos.2022.diversity}. Empirical studies on how humans perceive diversity in participant rankings have been carried out to validate the process.


\section{Diversity as an Instrumental Value that Supports Fundamental Values}
\label{S6}

Any attempt to operationalize and mediate diversity in social interactions must consider the possible ethical ramifications. Technological mediation is not value-neutral, and the development of technology is never done without prioritizing different values and advancing a certain ``agenda'' \cite{Friedman.1996,Brey.2010,schelenz2024transparency}. Being mindful of the inevitably value-ridden nature of the technology development process, our platform --- born out of a European, democratic, and liberal context --- seeks to promote the participants' exposure to diversity to the extent that it does not pose a threat to their safety. We, thus, have to carefully weigh the values of inclusion and protection to enable a meaningful exposure to diversity. 

Our rationale follows the assumption that diversity is not an end, but a means; it has instrumental rather than intrinsic value \cite{Zimmerman.2019}. In the same way that diversity may promote the value of inclusiveness, it may also, if abused, demote the value of protection. This relates to the classic discussion in content moderation: the need to balance the freedom of speech (as promoted in liberal democracies) and the prevention of hate speech and violence \cite{Gillespie.2018,Brown.26February2021}. On the one hand, public discourse should include as many perspectives as possible. On the other hand, public discourse should be free from abusive communication (such as racist, sexist, and antisemitic attacks) and protect minority positions. 

Values like inclusiveness and protection are of a more fundamental nature vis \`{a} vis diversity \cite{Helm.23.09.2021,Helm.2022}, and they ought to be promoted for their own sake, even at the expense of diversity. Think of a request about a topic that is associated with stigma (e.g., mental health problems), where an individual may risk social exclusion just for seeking support \cite{Goffman.1990}. Somewhat counter-intuitively, restricting the diversity of the respondents ends up enhancing the individual’s access to a diverse section of the social search space, by protecting them from hurtful and derogatory comments, and by creating a safe space within which to seek support. Accordingly, our platform implements algorithms and offers interface functionalities to interact anonymously and to restrict the  respondent pool. 


\section{Conclusion}

We have sought to make progress on an ambitious goal: to design an AI-mediated and diversity-aware platform for social interactions. In attempting to fulfill that goal, we have explored various facets of diversity, while, importantly, staying grounded in empirical work across a diverse set of participating communities. Among our contributions is the gathering, and the eventual sharing, of social interaction data at a scale that goes beyond past attempts. This is only a first step, and there is still a long way to go. But we believe that the direction of rethinking social interactions under a diversity paradigm is promising, and it does raise several relevant research questions. Ultimately, the main take-home message is the importance of a holistic, empirically-grounded, and multidisciplinary perspective in pushing towards a future of diversity-aware hybrid human-AI societies.
 

\section*{Acknowledgements}

This research has received funding from the European Union's Horizon 2020 FET Proactive project ``WeNet --- The Internet of Us'', under grant agreement no.\ 823783. We wish to thank all our collaborators for their amazing support; a, most likely, incomplete list is: Marina Bidoglia, Andrea Bontempelli, Maria Chiara Campodonico, Carlo Caprini, Ronald Chenu Abente, Galileo Disperati, Paula Helm, Alethia Hume, Peter Kun, Meegahapaola Lakmal, Vasileios Theodoros Markos, Marcelo Dario Rodas Brites, and Donglei Song.


\bibliographystyle{vancouver}
\bibliography{references}

@inproceedings{botta2023sequencing,
 title={Sequencing Educational Content Using Diversity Aware Bandits.},
author={Botta, Colton and Segal, Avi and Gal, Kobi},
booktitle={EDM},
year={2023}
}

@article{botta2023personalizing,
 title={Personalizing Interventions with Diversity Aware Bandits},
author={Botta, Colton and Segal, Avi and Gal, Kobi},
journal={HHAI-WS 2023},
year={2023}
}

@article{schelenz2024transparency,
title={Transparency-Check: An Instrument for the Study and Design of Transparency in AI-based Personalization Systems},
author={Schelenz, Laura and Segal, Avi and Adelio, Oduma and Gal, Kobi},
journal={ACM Journal on Responsible Computing},
volume={1},
number={1},
pages={1--18},
year={2024},
publisher={ACM New York, NY}
}

@Article{Tsui1991,
  author  = {Tsui, A.S. and Terri, Egan and O'Reilly, C.},
  journal = {Academy of Management Proceedings},
  title   = {{Being Different: Relational Demography and Organizational Attachment}},
  year    = {1991},
  pages   = {183--187},
  volume  = {1},
}

@Article{Tsui1992,
  author  = {Tsui, A.S. and Terri, Egan and O'Reilly, C.},
  journal = {Administrative Science Quarterly},
  title   = {{Being Different: Relational Demography and Organizational Attachment}},
  year    = {1992},
  pages   = {549--579},
  volume  = {37(4)},
}

@Article{Harrison1998,
  author  = {Harrison, David. A. and Price, K. H. and Bell, M. P.},
  journal = {Academy of Management Journal},
  title   = {{Beyond Relational Demography: Time and the Effects of Surface- and Deep-Level Diversity on Work Group Cohesion}},
  year    = {1998},
  pages   = {96--107},
  volume  = {41(1)},
}

@Article{Rapke2009,
  author  = {R{\o}pke, I.},
  journal = {Ecological Economics},
  title   = {{Theories of Practice --- New Inspiration for Ecological Economic Studies on Consumption}},
  year    = {2009},
  pages   = {2490--2497},
  volume  = {68(10)},
}

@Article{shove2005,
  author  = {Shove, E. and Pantzar, M.},
  journal = {Journal of Consumer Culture},
  title   = {{Consumers, Producers and Practices: Understanding the Invention and Reinvention of Nordic Walking}},
  year    = {2005},
  pages   = {43--64},
  volume  = {5(1)},
}

@Book{berger1996,
  author    = {Berger, P. and Luckmann, T.},
  publisher = {Garden City, NJ: Doubleday.},
  title     = {{The Social Construction of Reality}},
  year      = {1966},
}

@InProceedings{KD-2020-Osman,
  author    = {Osman, Nardine and Sierra, Carles and Chenu-Abente, Ronald and Qiang, Shen and Giunchiglia, Fausto},
  booktitle = {17th European Conference on Multi-Agent Systems (EUMAS)},
  title     = {{Open Social Systems}},
  year      = {2020},
  address   = {Thessaloniki, Greece},
}

@Misc{Helm.23.09.2021,
  author   = {Helm, Paula and Michael, Loizos and Schelenz, Laura},
  title    = {{Diversity by Design: Balancing Protection and Inclusion in Social Networks}},
  date     = {2021-09-23},
  url      = {https://arxiv.org/pdf/2109.11484},
}

@InProceedings{Helm.2022,
    author={Paula Helm and Loizos Michael and Laura Schelenz},
    title={{``Diversity by Design''? Balancing the Inclusion and Protection of Users in an Online Social Platform}},
    booktitle={Proceedings of the 2022 AAAI/ACM Conference on Artificial Intelligence, Ethics and Society (AIES '22)},
    pages={324--334},
    year={2022},
    url={https://doi.org/10.1145/3514094.3534149}
    }

@inproceedings{bella2024tackling,
  title={Tackling Language Modelling Bias in Support of Linguistic Diversity},
  author={Bella, G{\'a}bor and Helm, Paula and Koch, Gertraud and Giunchiglia, Fausto},
  booktitle={The 2024 ACM Conference on Fairness, Accountability, and Transparency},
  pages={562--572},
  year={2024}
}

@article{giunchiglia2023representing,
  title={Representing interlingual meaning in lexical databases},
  author={Giunchiglia, Fausto and Bella, G{\'a}bor and Nair, Nandu C and Chi, Yang and Xu, Hao},
  journal={Artificial Intelligence Review},
  volume={56},
  number={10},
  pages={11053--11069},
  year={2023},
  publisher={Springer}
}

@article{giunchiglia2022representation,
  title={Representation heterogeneity},
  author={Giunchiglia, Fausto and Bagchi, Mayukh},
  journal={arXiv preprint arXiv:2207.01091},
  year={2022}
}

@article{bontempelli2022lifelong,
  title={Lifelong Personal Context Recognition},
  author={Bontempelli, Andrea and Britez, Marcelo Rodas and Li, Xiaoyue and Zhao, Haonan and Erculiani, Luca and Teso, Stefano and Passerini, Andrea and Giunchiglia, Fausto},
  journal={arXiv preprint arXiv:2205.10123},
  year={2022}
}

@article{mercado2023social,
  title={Social interactions mediated by the Internet and the Big-Five: a cross-country analysis},
  author={Mercado, Andrea and Hume, Alethia and Bison, Ivanno and Giunchiglia, Fausto and Ganbold, Amarsanaa and Cernuzzi, Luca},
  journal={arXiv preprint arXiv:2309.10681},
  year={2023}
}

@article{girardini2023adaptation,
  title={Adaptation of student behavioural routines during Covid-19: a multimodal approach},
  author={Girardini, Nicol{\`o} Alessandro and Centellegher, Simone and Passerini, Andrea and Bison, Ivano and Giunchiglia, Fausto and Lepri, Bruno},
  journal={EPJ Data Science},
  volume={12},
  number={1},
  pages={55},
  year={2023},
  publisher={Springer Berlin Heidelberg}
}

@article{koch2024layers,
  title={Layers of technology in pluriversal design decolonising language technology with the live language initiative},
  author={Koch, Gertraud and Bella, G{\'a}bor and Helm, Paula and Giunchiglia, Fausto},
  journal={CoDesign},
  volume={20},
  number={1},
  pages={77--90},
  year={2024},
  publisher={Taylor \& Francis}
}

@InProceedings{UKC-IJCAI,
  author    = {Giunchiglia, Fausto and Batsuren, Khuyagbaatar and Bella, Gabor},
  booktitle = {IJCAI},
  title     = {{Understanding and Exploiting Language Diversity}},
  year      = {2017},
  pages     = {4009--4017},
}

@InProceedings{KD-2020-Giunchiglia-KR,
  author    = {Giunchiglia, Fausto and Fumagalli, Mattia},
  booktitle = {Knowledge Representation Conference (KR)},
  title     = {{Entity Type Recognition -- Dealing with the Diversity of Knowledge}},
  year      = {2020},
  address   = {Rhodes, Greece},
}

@Article{orphanou2022mitigating,
  author    = {Orphanou, Kalia and Otterbacher, Jahna and Kleanthous, Styliani and Batsuren, Khuyagbaatar and Giunchiglia, Fausto and Bogina, Veronika and Tal, Avital Shulner and Hartman, Alan and Kuflik, Tsvi},
  journal   = {ACM Computing Surveys},
  title     = {{Mitigating Bias in Algorithmic Systems --- A Fish-Eye View}},
  year      = {2022},
  number    = {5},
  pages     = {1--37},
  volume    = {55},
  publisher = {ACM New York, NY},
}

@Misc{KD-2022-WeNetDiversityOnePlus,
  author       = {Giunchiglia, Fausto and Bison, Ivano and Busso, Matteo and Chenu-Abente, Ronald and Rodas, Marcelo and Zeni, Mattia and Can, Gunel and Giuseppe, Veltri and Amalia, DeG{\"o}tzen and Peter, Kun and Amarsanaa, Ganbold and Altangerel, Chagnaa and George, Gaskell and Sally, Stares and Miriam, Bidoglia and Luca, Cernuzzi and Alethia, Hume and Jose Luis, Zarza and Hao, Xu and Donglei, Song and Shyam, Diwakar and Chaitanya, Nutakki and Salvador, Ruiz Correa and Andrea-Rebeca, Mendoza and Lakmal, Meegahapola and Daniel, Gatica-Perez},
  howpublished = {See \url{https://datascientia.disi.unitn.it/}.},
  note         = {University of Trento Technical Report - DataScientia dataset descriptors},
  title        = {{A Worldwide Diversity Pilot on Daily Routines and Social Practices (2020-2021)}},
  year         = {2022},
}

@InProceedings{KD-2017-PERCOM,
  author       = {Giunchiglia, Fausto and Bignotti, Enrico and Zeni, Mattia},
  booktitle    = {2017 IEEE International Conference on Pervasive Computing and Communications Workshops (PerCom Workshops)},
  title        = {{Personal Context Modelling and Annotation}},
  year         = {2017},
  organization = {IEEE},
  pages        = {117--122},
}

@inproceedings{zeni2014multi,
  title={Multi-device activity logging},
  author={Zeni, Mattia and Zaihrayeu, Ilya and Giunchiglia, Fausto},
  booktitle={Proc. of the  ACM PERCOM conference - adjunct publication},
  pages={299--302},
  year={2014}
}

@InProceedings{schelenz2021theory,
  author    = {Schelenz, Laura and Bison, Ivano and Busso, Matteo and De G{\"o}tzen, Amalia and Gatica-Perez, Daniel and Giunchiglia, Fausto and Meegahapola, Lakmal and Ruiz-Correa, Salvador},
  booktitle = {Proceedings of the 2021 AAAI/ACM Conference on AI, Ethics, and Society},
  title     = {{The Theory, Practice, and Ethical Challenges of Designing a Diversity-Aware Platform for Social Relations}},
  year      = {2021},
  pages     = {905--915},
}

@Article{Friedman.1996,
  author  = {Friedman, Batya and Nissenbaum, Helen},
  journal = {ACM Transactions on Information Systems},
  title   = {{Bias in Computer Systems}},
  year    = {1996},
  issn    = {10468188},
  number  = {3},
  pages   = {330--347},
  volume  = {14},
  doi     = {10.1145/230538.230561},
}

@InCollection{Brey.2010,
  author    = {Brey, Philip},
  booktitle = {The Cambridge Handbook of Information and Computer Ethics},
  publisher = {{Cambridge University Press}},
  title     = {{Values in Technology and Disclosive Computer Ethics}},
  year      = {2010},
  address   = {Cambridge},
  editor    = {Floridi, Luciano},
  isbn      = {0511845235},
  pages     = {41--58},
  doi       = {10.1017/CBO9780511845239.004},
}

@Book{Gillespie.2018,
  author    = {Gillespie, Tarleton},
  publisher = {{Yale University Press}},
  title     = {{Custodians of the Internet: Platforms, Content Moderation, and the Hidden Decisions that Shape Social Media}},
  year      = {2018},
  address   = {New Haven},
  isbn      = {9780300173130},
}

@Misc{Brown.26February2021,
  author       = {Brown, Alexander},
  title        = {{What is Hate Speech? Presentation of Hard Cases}},
  year         = {26 February 2021},
  address      = {Virtual},
  institutions = {{Karlsruhe Institute of Technology}},
  series       = {Workshop Hate Speech: What It Is and How It Works},
}

@Book{Goffman.1990,
  author    = {Goffman, Erving},
  publisher = {Penguin},
  title     = {{Stigma: Notes on the Management of Spoiled Identity}},
  year      = {1990},
  address   = {London},
  isbn      = {978-0140124750},
  series    = {Penguin Psychology},
  price     = {{\pounds} 4.99},
}

@Misc{Zimmerman.2019,
  author  = {Zimmerman, Michael J. and Bradley, Ben},
  title   = {{Intrinsic vs. Extrinsic Value}},
  year    = {2019},
  url     = {https://plato.stanford.edu/entries/value-intrinsic-extrinsic/},
  urldate = {23 July 2021},
}

@InProceedings{yanovsky2019one,
  author    = {Yanovsky, Stav and Hoernle, Nicholas and Lev, Omer and Gal, Kobi},
  booktitle = {Proceedings of the 27th ACM Conference on User Modeling, Adaptation and Personalization},
  title     = {{One Size Does Not Fit All: Badge Behavior in Q\&A Sites}},
  year      = {2019},
  pages     = {113--120},
}

@InProceedings{segal2018optimizing,
  author    = {Segal, Avi and Gal, Kobi and Kamar, Ece and Horvitz, Eric and Miller, Grant},
  booktitle = {Proceedings of the AAAI Conference on Artificial Intelligence},
  title     = {{Optimizing Interventions via Offline Policy Evaluation: Studies in Citizen Science}},
  year      = {2018},
  number    = {1},
  volume    = {32},
}

@InProceedings{ben2022generating,
  author    = {Ben Zaken, Daniel and Segal, Avi and Cavalier, Darlene and Shani, Guy and Gal, Kobi},
  booktitle = {Proceedings of the 30th ACM Conference on User Modeling, Adaptation and Personalization},
  title     = {{Generating Recommendations with Post-Hoc Explanations for Citizen Science}},
  year      = {2022},
  pages     = {69--78},
}

@Article{10.1145/3569483,
  author     = {Meegahapola, Lakmal and Droz, William and Kun, Peter and de G\"{o}tzen, Amalia and Nutakki, Chaitanya and Diwakar, Shyam and Correa, Salvador Ruiz and Song, Donglei and Xu, Hao and Bidoglia, Miriam and Gaskell, George and Chagnaa, Altangerel and Ganbold, Amarsanaa and Zundui, Tsolmon and Caprini, Carlo and Miorandi, Daniele and Hume, Alethia and Zarza, Jose Luis and Cernuzzi, Luca and Bison, Ivano and Britez, Marcelo Rodas and Busso, Matteo and Chenu-Abente, Ronald and G\"{u}nel, Can and Giunchiglia, Fausto and Schelenz, Laura and Gatica-Perez, Daniel},
  journal    = {Proceedings of the ACM on Interactive, Mobile, Wearable and Ubiquitous Technologies},
  title      = {{Generalization and Personalization of Mobile Sensing-Based Mood Inference Models: An Analysis of College Students in Eight Countries}},
  year       = {2023},
  number     = {4},
  volume     = {6},
  address    = {New York, NY, USA},
  doi        = {10.1145/3569483},
  issue_date = {December 2022},
  publisher  = {Association for Computing Machinery},
  url        = {https://doi.org/10.1145/3569483},
}

@InProceedings{Bouton22,
  author    = {Bouton-Bessac, Emma and Gatica-Perez, Daniel and Meegahapola, Lakmal Buddika},
  booktitle = {in Proc. Int. Conf. on Pervasive Computing Technologies for Healthcare (Pervasive Health)},
  title     = {{Your Day in Your Pocket: Complex Activity Recognition from Smartphone Accelerometers}},
  year      = {2022},
  month     = dec,
}

@InProceedings{GaticaPerez19,
  author    = {Gatica-Perez, Daniel and Santani, Darshan and Biel, Joan-Isaac and and Phan, Trung},
  booktitle = {in Proc. ACM Workshop on Fairness, Accountability, and Transparency in Multimedia (FAT/MM)},
  title     = {{Social Multimedia, Diversity, and Global South Cities: A Double Blind Side}},
  year      = {2019},
  month     = oct,
}

@Article{10.1145/3458723,
  author     = {Gebru, Timnit and Morgenstern, Jamie and Vecchione, Briana and Vaughan, Jennifer Wortman and Wallach, Hanna and III, Hal Daum\'{e} and Crawford, Kate},
  journal    = {Commun. ACM},
  title      = {{Datasheets for Datasets}},
  year       = {2021},
  issn       = {0001-0782},
  month      = nov,
  number     = {12},
  pages      = {86--92},
  volume     = {64},
  address    = {New York, NY, USA},
  doi        = {10.1145/3458723},
  issue_date = {December 2021},
  numpages   = {7},
  publisher  = {Association for Computing Machinery},
  url        = {https://doi.org/10.1145/3458723},
}

@Article{cinelli2021echo,
  author    = {Cinelli, Matteo and De Francisci Morales, Gianmarco and Galeazzi, Alessandro and Quattrociocchi, Walter and Starnini, Michele},
  journal   = {Proceedings of the National Academy of Sciences},
  title     = {{The Echo Chamber Effect on Social Media}},
  year      = {2021},
  number    = {9},
  pages     = {e2023301118},
  volume    = {118},
  publisher = {National Acad Sciences},
}

@InProceedings{Hutchinson,
  author    = {Hutchinson, Hilary and Mackay, Wendy and Westerlund, Bo and Bederson, Benjamin B. and Druin, Allison and Plaisant, Catherine and Beaudouin-Lafon, Michel and Conversy, St\'{e}phane and Evans, Helen and Hansen, Heiko and Roussel, Nicolas and Eiderb\"{a}ck, Bj\"{o}rn},
  booktitle = {Proceedings of the SIGCHI Conference on Human Factors in Computing Systems},
  title     = {{Technology Probes: Inspiring Design for and with Families}},
  year      = {2003},
  address   = {New York, NY, USA},
  pages     = {17--24},
  publisher = {Association for Computing Machinery},
  series    = {CHI '03},
  doi       = {10.1145/642611.642616},
  isbn      = {1581136307},
  location  = {Ft. Lauderdale, Florida, USA},
  numpages  = {8},
  url       = {https://doi.org/10.1145/642611.642616},
}

@InProceedings{Kun,
  author    = {Kun, Peter and de G\"otzen, Amalia and Bidoglia, Miriam and Gommesen, Niels and Gaskell, George},
  booktitle = {in Proc. DRS022: Bilbao},
  title     = {{Exploring Diversity Perceptions in a Community Through a Q\&A Chatbot}},
  year      = {2022},
  month     = jun,
}

@InProceedings{SantosOS22,
  author    = {Thiago Freitas dos Santos and Nardine Osman and Marco Schorlemmer},
  booktitle = {21st International Conference on Autonomous Agents and Multiagent Systems, {AAMAS} 2022, Auckland, New Zealand, May 9-13, 2022},
  title     = {{Ensemble and Incremental Learning for Norm Violation Detection}},
  year      = {2022},
  editor    = {Piotr Faliszewski and Viviana Mascardi and Catherine Pelachaud and Matthew E. Taylor},
  pages     = {427--435},
  publisher = {International Foundation for Autonomous Agents and Multiagent Systems {(IFAAMAS)}},
  bibsource = {dblp computer science bibliography, https://dblp.org},
  doi       = {10.5555/3535850.3535899},
  url       = {https://www.ifaamas.org/Proceedings/aamas2022/pdfs/p427.pdf},
}

@Article{OsmanCSSG21,
  author    = {Nardine Osman and Ronald Chenu{-}Abente and Qiang Shen and Carles Sierra and Fausto Giunchiglia},
  journal   = {{SN} Comput. Sci.},
  title     = {{Empowering Users in Online Open Communities}},
  year      = {2021},
  number    = {4},
  pages     = {338},
  volume    = {2},
  bibsource = {dblp computer science bibliography, https://dblp.org},
  doi       = {10.1007/s42979-021-00714-5},
  url       = {https://doi.org/10.1007/s42979-021-00714-5},
}

@Article{abs-2110-09240,
  author     = {Carles Sierra and Nardine Osman and Pablo Noriega and Jordi Sabater{-}Mir and Antoni Perell{\'{o}}},
  journal    = {CoRR},
  title      = {{Value Alignment: A Formal Approach}},
  year       = {2021},
  volume     = {abs/2110.09240},
  bibsource  = {dblp computer science bibliography, https://dblp.org},
  eprint     = {2110.09240},
  eprinttype = {arXiv},
  url        = {https://arxiv.org/abs/2110.09240},
}

@Article{MontesOS22,
  author    = {Nieves Montes and Nardine Osman and Carles Sierra},
  journal   = {Artif. Intell.},
  title     = {{A Computational Model of Ostrom's Institutional Analysis and Development Framework}},
  year      = {2022},
  pages     = {103756},
  volume    = {311},
  bibsource = {dblp computer science bibliography, https://dblp.org},
  doi       = {10.1016/j.artint.2022.103756},
  url       = {https://doi.org/10.1016/j.artint.2022.103756},
}

@InProceedings{MontesS21,
  author    = {Nieves Montes and Carles Sierra},
  booktitle = {{AAMAS} '21: 20th International Conference on Autonomous Agents and Multiagent Systems, Virtual Event, United Kingdom, May 3-7, 2021},
  title     = {{Value-Guided Synthesis of Parametric Normative Systems}},
  year      = {2021},
  editor    = {Frank Dignum and Alessio Lomuscio and Ulle Endriss and Ann Now{\'{e}}},
  pages     = {907--915},
  publisher = {{ACM}},
  bibsource = {dblp computer science bibliography, https://dblp.org},
  doi       = {10.5555/3463952.3464060},
  url       = {https://www.ifaamas.org/Proceedings/aamas2021/pdfs/p907.pdf},
}

@Article{Garcia-CaminoRSV09,
  author    = {Andr{\'{e}}s Garc{\'{\i}}a{-}Camino and Juan A. Rodr{\'{\i}}guez{-}Aguilar and Carles Sierra and Wamberto Weber Vasconcelos},
  journal   = {Auton. Agents Multi Agent Syst.},
  title     = {{Constraint Rule-Based Programming of Norms for Electronic Institutions}},
  year      = {2009},
  number    = {1},
  pages     = {186--217},
  volume    = {18},
  bibsource = {dblp computer science bibliography, https://dblp.org},
  doi       = {10.1007/s10458-008-9059-4},
  url       = {https://doi.org/10.1007/s10458-008-9059-4},
}

@Article{AgotnesHRSW09,
  author    = {Thomas {\AA}gotnes and Wiebe van der Hoek and Juan A. Rodr{\'{\i}}guez{-}Aguilar and Carles Sierra and Michael J. Wooldridge},
  journal   = {Stud Logica},
  title     = {{Multi-Modal {CTL:} Completeness, Complexity, and an Application}},
  year      = {2009},
  number    = {1},
  pages     = {1--26},
  volume    = {92},
  bibsource = {dblp computer science bibliography, https://dblp.org},
  doi       = {10.1007/s11225-009-9184-3},
  url       = {https://doi.org/10.1007/s11225-009-9184-3},
}

@InCollection{KowalskiS05,
  author    = {Robert A. Kowalski and Marek J. Sergot},
  booktitle = {The Language of Time - {A} Reader},
  publisher = {Oxford University Press},
  title     = {{A Logic-based Calculus of Events}},
  year      = {2005},
  editor    = {Inderjeet Mani and James Pustejovsky and Robert J. Gaizauskas},
  pages     = {217--240},
  bibsource = {dblp computer science bibliography, https://dblp.org},
}

@InProceedings{Perikos.2022,
    author    = {Isidoros Perikos and Loizos Michael}, 
    title     = {{A Survey on Tie Strength Estimation Methods in Online Social Networks}},
    booktitle = {Proceedings of the 14th International Conference on Agents and Artificial Intelligence - ICAART2022},
    pages     = {484--491},
    doi       = {10.5220/0010845100003116}
    }

@InProceedings{Markos.2022.prudens,
  author    = {Vasileios Theodoros Markos and Loizos Michael},
  booktitle = {Proceedings of the 6th International Joint Conference on Rules and Reasoning (RuleML+RR'22)},
  title     = {{Prudens: An Argumentation-Based Language for Cognitive Assistants}},
  year      = {2022},
  address   = {virtual / Berlin, Germany},
}

@Article{Dietz.2022.calculus,
  author    = {Emmanuelle Dietz and Antonis Kakas and Loizos Michael},
  title     = {{Argumentation: A Calculus for Human-Centric AI}},
  journal   = {Frontiers in Artificial Intelligence},
  volume    = {5},
  year      = {2022},
}

@InProceedings{Michael.2019.coaching,
  author    = {Loizos Michael},
  booktitle = {Proceedings of the IJCAI 2019 Workshop on Explainable Artificial Intelligence (XAI'19)},
  title     = {{Machine Coaching}},
  year      = {2019},
  address   = {S.A.R. Macau, P.R. China},
}

@InCollection{Michael.2023.coachable,
  author    = {Loizos Michael},
  title     = {{Coachable AI}},
  booktitle = {{ERCIM News No. 132, Special Theme: Cognitive AI \& Cobots}},
  year      = {2023},
}

@InProceedings{Markos.2022.proxy,
  author    = {Vasileios Theodoros Markos and Marios Thoma and Loizos Michael},
  booktitle = {Proceedings of the COMMA 2022 International Workshop on Argumentation and Machine Learning (ArgML'22)},
  title     = {{Machine Coaching with Proxy Coaches}},
  year      = {2022},
  address   = {Cardiff, Wales, U.K.},
}

@InProceedings{Ioannou.2021.translation,
  author    = {Christodoulos Ioannou and Loizos Michael},
  title     = {{Knowledge-Based Translation of Natural Language into Formal Logic}},
  booktitle = {7th IJCAI Workshop on Linguistic and Cognitive Approaches to Dialog
	Agents (LaCATODA)},
  year      = {2021},
}

@InProceedings{Tsamoura.2021,
  author    = {Efthymia Tsamoura and Timothy Hospedales and Loizos Michael},
  booktitle = {Proceedings of the 35th AAAI Conference on Artificial Intelligence (AAAI'21)},
  title     = {{Neural-Symbolic Integration: A Compositional Perspective}},
  year      = {2021},
  address   = {virtual},
}

@InCollection{Michael.2023.fourthAI,
  author    = {Loizos Michael},
  title     = {{Explainability and the Fourth AI Revolution}},
  booktitle = {{Handbook of Research on Artificial Intelligence, Innovation and Entrepreneurship}},
  editor    = {Elias G. Carayannis and Evangelos Grigoroudis},
  publisher = {Edward Elgar Publishing},
  year      = {2023},
}

@InProceedings{Tubella.2020,
  author    = {Andrea Aler Tubella and Andreas Theodorou and Virginia Dignum and Loizos Michael},
  booktitle = {Proceedings of the 4th International Joint Conference on Rules and Reasoning (RuleML+RR'20)},
  title     = {{Contestable Black Boxes}},
  year      = {2020},
  address   = {Oslo, Norway},
}

@InProceedings{Markos.2022.diversity,
    author    = {Vassilis Markos and Loizos Michael}, 
    title     = {{Post-Hoc Diversity-Aware Curation of Rankings}},
    booktitle = {Proceedings of the 14th International Conference on Agents and Artificial Intelligence (ICAART'22)},
    address   = {virtual},
    year      = {2022},
}

\end{document}